# Graphical Visualization of Risk Assessment for Effective Risk Management during Software Development Process

**Raghavi K Bhujang [1] & Suma. V[2]**

[1]Dept of MCA, PESIT, Bangalore & Member, Research and Industry Incubation Centre, RIIC, DSI, Bangalore
[2]Research and Industry Incubation Center, Dayananda Sagar Institutions, Bangalore, India
E-mail : raghavi@pes.edu, sumavdsce@gmail.com

*Abstract* - Success of any IT industry depends on the success rate of their projects, which in turn depends on several factors such as cost, time, and availability of resources. These factors formulate the risk areas, which needs to be addressed in a proactive way. The rudimentary objective of risk management is to circumvent the possibility of their occurrence by identifying the risks, preparing the contingency plans and mitigation plans in order to reduce the consequences of the risks. Hence, effective risk management becomes one of the imperative challenges in any organization, which if deemed in an apt way assures the continued sustainability of the organization in the high-end competitive environment. This paper provides visualization of risk assessment through a graphical model. Further, the matrix representation of the risk assessment aids the project personnel to identify all the risks, comprehend their frequency and probability of their occurrence. In addition, the graphical model enables one to analyze the impact of identified risks and henceforth to assign their priorities. This mode of representation of risk assessment factors helps the organization in accurate prediction of success rate of the project.

## I. INTRODUCTION

The survivability of any industry is driven by the successful nature of projects that are engineered on board. However, the success of any project depends on several influencing areas, which require consistent and periodical attention by the project manager. Some of these areas are schedule management, finance management, change management, conflict management, etc. It is worth to note that each of the aforementioned areas emerges as risk if not managed in righteous way. Risk is a chance for occurrence of an unlikely event, which would result with highly unacceptable consequences. Thus, risk management is one of the critical challenges that need to be addressed in a skillful and efficient manner. It is therefore necessary to take preventive measures to reduce the likelihood of risk occurrences and to minimize its impact in realizing effective project management.

The process of risk management starts with identification of risks and classifying them into different types. Subsequently, every classified risk is assessed with impact level, its probability of occurrence and frequency of occurrence that enables one to prioritize according to their severity. The systematic analysis of risk aids project personnel in achieving an accurate predictive estimation of the apt choice of process and resources in the project.

State of the art in the risk management domain indicates progress of research in risk associated issues. Authors of [1] recommend the classification of every identified risk in a risk based tree structure in order to assess and resolve them efficiently. They suggest probabilistic calculation approach for qualitative and quantitative analysis and assessment of these risks. However, classification of risk should include not only the type and it is essential to consider the probability and frequency of occurrence of every identified risk. This facilitates to realize the impact of the risks, which results in formulation of prioritization list.

This paper aims to bring in a crispy classification of risk based on several modulating factors such as risk type, risk probability and risk frequency, which are deemed three independent factors. Consequently, establishing dependency relation between these factors assists to analyze the risk impact efficiently using matrix representation.

The organization of this paper is as follows, section II of this paper provides details of the related work in the domain of risk management. Section III elucidates the analysis of the investigation. Section IV provides matrix representation and mathematical analysis of the risk assessment. The last section of this paper provides the summery of this investigation.

## II. LITERATURE SURVEY

Since the evolution of software engineering, risk management has become one of the key challenges in day-to-day software development activities. Authors of [2] have introduced a systematic and qualitative project analysis of risk using Risk Factor Analysis (RFA) method. They state that RFA technique enables risk analysts to develop projects effectively.

Study made by authors of [3] on data collected over several industries indicates the extent of usage of risk management practices in industries. They suggest the impact of risk management to depend mainly on better meeting time and budget goals and less on product performance and specification. However, authors of [4] focus on the benefits of implementing effective risk management tools and techniques in software development project.

Further, authors in [5] analyzed the risks involved during software design. They feel that business goals determine risks and risk drives methods while methods yield measurement resulting in measurement to further drive decision support. Consequently, they state that decision support drives fix/rework and application quality.

On the other hand, authors of [6] have developed a tool named Risk Failure Mode and Effects Analysis (RFMEA) which is an extension of the Failure Mode and Effects Analysis technique. The benefit of the RFMEA include an increased focus on the most imminent risks, prioritizing risk contingency planning, improved team participation in the risk management process, and development of improved risk controls.

Authors of [7] have investigated the various risk sources in design and build projects quality using the discriminant analysis technique where risk is classified into one of the three risk groups namely cost, time and quality. However, authors of [8] provide an explanation on what enterprise risk management is and how an operational risk management fits into the ERM framework.

## III. RISK ASSESSMENT

Risk is the consequence of inadequate information. Every identified risk is assessed for its classification. Classification of risks further enables one to evaluate the vitality of the risk in the effective development process. It is further elucidated below reflecting the significant activities to be formulated in order to develop high quality risk-free deliverable.

*A. Risk Identification*

Identifying the risks is not a component of a single area or a phase in software development and is a requisite activity involved in the entire project. This process involves all personnel who will be affected by the risks such as customers, stakeholders, Subject Matter Experts (SME), Risk management experts, project managers, team members and end users. Process areas, which fall into the uncertainty level causes threat and hence deemed as risk areas in project management. Thus, all the uncertainty areas need to be focussed to identify the risk factors, which are either internal or external risk factors and are estimated from the development level up to the organisational level. This process of risk identification begins at the project conceiving point, up to the project deployment point as risk can crop up at any point of time and in any phase within a project, which needs to be monitored periodically.

In lieu of avoiding the risks from popping up, risks need to be welcomed within a project by which a mitigation plan can be prepared which lead to the reduction of the level of uncertainty in future projects. With identification of every risk (irrespective of the phase of the project), it is categorized in to different risk types depending on the area that gets the impact and further based on the probability of occurrence of the risk in addition to consideration of frequency of its occurrence.

*B. Risk Classification*

With the assessment of every identified risk, it is a wise justification to further classify the risks. Our wide spectrum of investigation on several projects across various software industries has resulted in the classification of risk to be depending on risk type, on probability of risk occurrence, on frequency of occurrence of the risk, on risk impact and on risk priority.

TABLE I. depicts the sample risk type classification which is shown at a higher level of abstraction.

TABLE I : SAMPLE RISK TYPES AT THE COARSE LEVEL OF ABSTRACTION

| Risk Type | Risk Type Description |
|---|---|
| Technology Risk | The Software / Technology related to the project may have risks. |
| Cost Risk | Risk associated with the ability of the project to achieve the planned life-cycle costs |
| Schedule Risk | Risk associated with the adequacy of the time allotted for the planning, R & D, |

Graphical Visualization of Risk Assessment for Effective Risk Management during Software Development Process

| | |
|---|---|
| | facility design, construction, and startup operations. |
| Scope | Risk that comes when Requirements are ignored for the sake of technology. |
| People | Project lacks enough staff or those with the right skills. |
| Requirements | Project changes are managed poorly. |
| Estimation | The time required to develop the software is underestimated. |
| Tools | The code generated by CASE tools is inefficient. CASE tools cannot be integrated. |
| Organizational | The organization is restructured so that different management are responsible for the project. |

TABLE I. infers the existence of several types of risks at a coarse level of abstraction.

Risk probability is the probability of occurrence of risk that normally varies from 0% occurrence to 100% occurrence where the boundary values of 0 and 100 will never be met. TABLE II. illustrates the probable encounter of risk in the development process. As an instance, the probability of occurrence of risk is 0% indicates the non-occurrence of the risk. However, the probability of risk occurrence being 100% is a case of certainty and hence is not considered under the probability classification of risk. TABLE II. infers that any type of risk, varying between 0 % and 100% is only considered for risk mitigation and further processing.

TABLE II : PROBABILITY OF RISK OCCURRENCE

| Risk Type | Probability of occurrence |
|---|---|
| Breach of organizational and project standards | 0% |
| Defects | 100% |
| All risks types as mentioned in Table 1. | 0<x<100 where x represents the percentage of probability of risk occurrence |

Since, not all risks can be avoided even after the implementation of a mitigation plan, the frequency of occurrence of these risks can take up different values before implementing this plan and after implementing the same. For instance, in a microwave application the performance of a timer is considered as a risk factor. Frequency of the timer not functioning may be 7 times in an hour and hence mitigation plan is formed to resolve the risk. With the implementation of the mitigation plan, the risk occurrence frequency would have come down to 2 times in an hour. TABLE III. illustrates the frequency of risk occurrence for the sample risk types.

TABLE III : FREQUENCY OF RISK OCCURRENCE FOR SAMPLE RISK TYPES

| Risk Type | Risk Frequency (probable frequency) | Risk Description |
|---|---|---|
| Requirements | Frequent | Likely to occur very often and/or continuously. |
| Schedule Risk | Likely | Occurs several times over the course of a transformation cycle. |
| People | Occasional | Occurs sporadically. |
| Technology Risk | Seldom | Remotely possible and would probably occur not more than once in the course of a transformation cycle. |
| Change in Organizational Standards | Unlikely | Will probably not occur during the course of a transformation cycle. |

TABLE III. infers a sample of risk types and its estimated frequency of occurrence. The frequency of occurrence of risks in project management is estimated using historical data regarding the number of times it has occurred until then and its subsequent occurrences.

IV. GRAPHICAL VISUALIZATION OF RISK CLASSIFICATION

Risk management is one of the core needs of the day. From this investigation, it is apparent that analysis of risk is influenced by several factors such as risk type, probability of risk occurrence, frequency of occurrence of specified risk. However, our study has thrown light on the fact that risk is independently influenced by the aforementioned modulating factors while the other risk influencing factors such as impact and priority of the risk are dependent on these three above-stated factors. Since the existence of dependency relations on risk influencing factors, it can be visualized in a mathematical model using graphical representation and matrix representational scheme. The mathematical visualization of risk factors enables one to accurately estimate and predict the efficiency of every deliverable during the software development process, which does currently not exist in the industrial atmosphere. Figure

1. depicts the graphical representation of risk assessment factors.

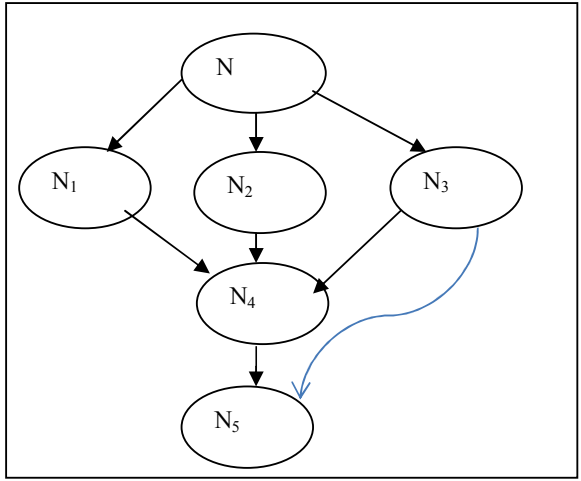

Fig. 1 : Graphical representation of risk assessment factors

Where, N – Risk

$N_1$ – Risk Type

$N_2$ – Risk Probability

$N_3$ – Risk Frequency

$N_4$ – Risk Impact

$N_5$ – Risk Priority

And ($N_1, N_2, N_3$) are the independent factors.

For any Risk (N), assessment of risk is made by considering its type, probability of occurrence and frequency ($N_1$, $N_2$, $N_3$). However, they further influence $N_4$, which in turn influences $N_5$. Hence, ($N_1$, $N_2$, $N_3$) influences N and influences $N_4$ while $N_4$ influences $N_5$. Thus, there exists a transitive relation, R, between the risk assessment factors. Figure 2. Figure 3. and Figure 4. represents aforementioned analysis.

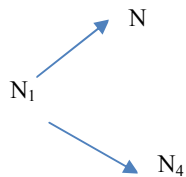

Fig. 2 : Assessment of Risk type

Figure 2 infers that R ($N_1$) = {N, $N_4$} i.e. Risk Type influences both Risk and the Risk Impact.

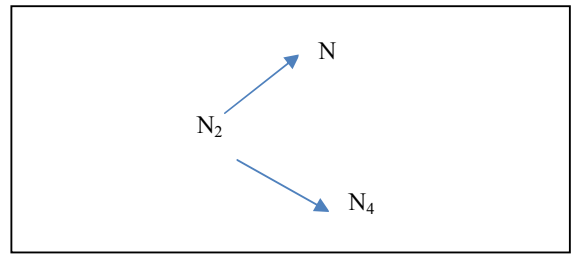

Fig. 3 : Assessment of Risk type

Figure 3 infers that R ($N_2$) = {N, $N_4$} i.e. Risk Probability also influences both Risk and the Risk Impact.

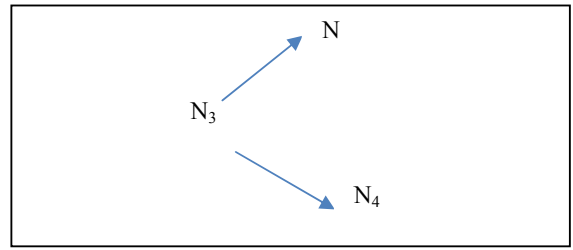

Fig. 4 : Assessment of Risk type

Figure 4 infers that R ($N_3$) = {N, $N_4$} i.e. Risk Frequency also influences both Risk and the Risk Impact. However,

N4 ⟶ N5 i.e. R($N_4$) = {$N_5$} i.e. Risk Impact influences Risk Priority.

Where these R ($N_1$)...R ($N_4$) are the relational sets and $N_1, N_2, N_3$ are the independent factors. These independent factors can take up different values depending on the type of project.

Figure 5. depicts a matrix representation of the above analysis.

$$M_R = \begin{pmatrix} & N & N1 & N2 & N3 & N4 & N5 \\ N & 0 & 0 & 0 & 0 & 0 & 0 \\ N1 & 1 & 0 & 0 & 0 & 0 & 0 \\ N2 & 1 & 0 & 0 & 0 & 0 & 0 \\ N3 & 1 & 0 & 0 & 0 & 0 & 0 \\ N4 & 0 & 0 & 0 & 0 & 0 & 1 \\ N5 & 0 & 0 & 0 & 0 & 0 & 0 \end{pmatrix}$$

Fig. 5 : Matrix representation of risk assessment factors

Figure 5. shows that $N_1 \rightarrow N_4 \rightarrow N_5$ i.e. $N_1$ (Risk Type) influences $N_4$ (Risk Impact) which in turn influences $N_5$ (Risk Priority).

This is a transitive relation which says that $N_1 \rightarrow N_5$., $N_2 \rightarrow N_5$ and also $N_3 \rightarrow N_5$

It is worth to note that, this can be further proved by applying FloydWarshall's Algorithm on matrix $M_R$. Floyd Warshall's algorithm is used for finding the transitive closure of a relation R. It takes as input the matrix $M_R$ representing the relation and output the matrix $M_{R*}$

$$M_R^* = \begin{pmatrix} N & N1 & N2 & N3 & N4 & N5 \\ N & 0 & 0 & 0 & 0 & 0 & 0 \\ N1 & 1 & 0 & 0 & 0 & 0 & 1 \\ N2 & 1 & 0 & 0 & 0 & 0 & 1 \\ N3 & 1 & 0 & 0 & 0 & 0 & 1 \\ N4 & 0 & 0 & 0 & 0 & 0 & 1 \\ N5 & 0 & 0 & 0 & 0 & 0 & 0 \end{pmatrix}$$

Where the values of column $N_5$ for $N_1$, $N_2$, $N_3$ are replaced by the new transitive values i.e.1.

This algorithm will be beneficial when the order of the matrix is high i.e. when more number of factors are involved which needs to be identified with new relationships.

The focus of this paper is to provide a graphical visualization of risk assessment factors and to provide a mathematical analysis of the relations existing between them. However, our forthcoming papers explore the impact analysis of these representations.

## V. CONCLUSION

Sustainability of any organization depends on the effective risk management capability of the company. Ever since the existence of industries, successful risk management has become one of the fundamental activities on board.

This investigation on several projects across various software industries has indicated that existing risk management operates on domain knowledge and experience of resource personnel. However, mathematical perspective of risk management is more result oriented than supposition oriented. This paper therefore focuses on graphical visualization of risk assessment, which enables the resource personnel to accurately estimate and predict the risk factors. Nevertheless, the mathematical representation is valid for known and identified risks.